\documentclass[sigconf, natbib=true, nonacm]{acmart}

\usepackage{hyperref}
\usepackage{makecell}

\newif\ifshowcomments
\showcommentstrue
\ifshowcomments
\newcommand{\mynote}[2]{\fbox{\bfseries\sffamily\scriptsize{#1}}
{\small$\blacktriangleright$\textsf{\emph{#2}}$\blacktriangleleft$}}
\else
\newcommand{\mynote}[2]{}
\fi

\AtBeginDocument{%
  \providecommand\BibTeX{{%
    \normalfont B\kern-0.5em{\scshape i\kern-0.25em b}\kern-0.8em\TeX}}}

\copyrightyear{2023}
\acmYear{2023}
\setcopyright{rightsretained}
\acmConference[UCC '23]{2023 IEEE/ACM 16th International Conference on Utility and Cloud Computing}{December 4--7, 2023}{Taormina (Messina), Italy}
\acmBooktitle{2023 IEEE/ACM 16th International Conference on Utility and Cloud Computing (UCC '23), December 4--7, 2023, Taormina (Messina), Italy}\acmDOI{10.1145/3603166.3632561}
\acmISBN{979-8-4007-0234-1/23/12}

\begin{document}


\title[Distributed Edge Inference: an Experimental Study on Multiview Detection]{Distributed Edge Inference:\\an Experimental Study on Multiview Detection}

\author{Gianluca Mittone}
\email{gianluca.mittone@unito.it}
\orcid{0000-0002-1887-6911}
\affiliation{%
  \institution{University of Turin}
  \streetaddress{Corso Svizzera 185}
  \city{Turin}
  \state{Piedmont}
  \country{Italy}
  \postcode{10149}
}

\author{Giulio Malenza}
\email{giulio.malenza@unito.it}
\orcid{0009-0006-4862-7429}
\affiliation{%
  \institution{University of Turin}
  \streetaddress{Corso Svizzera 185}
  \city{Turin}
  \state{Piedmont}
  \country{Italy}
  \postcode{10149}
}

\author{Marco Aldinucci}
\email{marco.aldinucci@unito.it}
\orcid{0000-0001-8788-0829}
\affiliation{%
  \institution{University of Turin}
  \streetaddress{Corso Svizzera 185}
  \city{Turin}
  \state{Piedmont}
  \country{Italy}
  \postcode{10149}
}

\author{Robert Birke}
\email{robert.birke@unito.it}
\orcid{0000-0003-1144-3707}
\affiliation{%
  \institution{University of Turin}
  \streetaddress{Corso Svizzera 185}
  \city{Turin}
  \state{Piedmont}
  \country{Italy}
  \postcode{10149}
}

\renewcommand{\shortauthors}{Mittone, et al.}

\begin{abstract}
Computing is evolving rapidly to cater to the increasing demand for sophisticated services, and Cloud computing lays a solid foundation for flexible on-demand provisioning. 
However, as the size of applications grows, the centralised client-server approach used by Cloud computing increasingly limits the applications' scalability.
To achieve ultra-scalability, cloud/edge/fog computing converges into the compute continuum, completely decentralising the infrastructure to encompass universal, pervasive resources.
The compute continuum makes devising applications benefitting from this complex environment a challenging research problem.
We put the opportunities the compute continuum offers to the test through a real-world multi-view detection model (MvDet) implemented with the FastFL C/C++ high-performance edge inference framework.
Computational performance is discussed considering many experimental scenarios, encompassing different edge computational capabilities and network bandwidths. 
We obtain up to 1.92x speedup in inference time over a centralised solution using the same devices.

\end{abstract}





\begin{center}
The following paper is the accepted version of ACM copyrighted material
\\[12pt]
\textit{Gianluca Mittone, Giulio Malenza, Marco Aldinucci, \& Robert Birke (2023). Distributed Edge Inference: an Experimental Study on Multiview Detection. In Proceedings of the IEEE/ACM 16th International Conference on Utility and Cloud Computing, UCC 2023, Taormina (Messina), Italy, December 4-7, 2023 (pp. 30). ACM.}
\\[12pt]
presented at the DML-ICC workshop at the UCC'23 conference in Taormina, Italy.
\\[12pt]
DOI: \href{https://doi.org/10.1145/3603166.3632561}{https://doi.org/10.1145/3603166.3632561}
\end{center}

\maketitle

\section{Introduction}
\label{sec:intro}

Recent machine learning techniques advancements lead to models reaching (and even surpassing) human capabilities on many tasks, fuelling a proliferation of AI-based applications. These applications mainly leverage the flexible computing and network resources cloud computing provides to achieve their scalability goals~\cite{you2019fast}. However, as the number of users and the complexity of deep neural network (DNN) models grows, the centralised approach taken by cloud computing is quickly reaching its limit. Especially AI-based applications with high resource costs per user request, such as transformer-based large language models, e.g. ChatGPT~\cite{RAY2023121}, and generative diffusion models, e.g. DallE~\cite{ramesh2022hierarchical}, put increasing strains on the cloud datacenters. New approaches try to alleviate this scalability bottleneck by gradually decentralising the infrastructure via ``walking-size'' cloud-connected server rooms pervasively distributed in the environment (edge computing) and the resources along the network paths (fog computing) up to a fully decentralised infrastructure (compute continuum)~\cite{kimovski2021compute}. The complexity stemming from such an environment's sheer size and heterogeneity makes designing applications able to take full benefits extremely challenging.

Traditionally, most ``smart'' applications leverage the classic client-server model where the user interacts with an app or a web API to send requests processed by (possibly clustered) servers in the cloud. On the one hand, the cloud can offer performance unreachable by smaller devices available in server rooms or even more at the edge. On the other hand, an increasing number of existing edge devices, like Raspberry Pi\cite{pahl2016raspberry} and most smart home devices~\cite{zhang2018home}, still offer enough computing power to handle inference of deep models, either unmodified or distilled~\cite{gou2021knowledge}.
However, most edge inference-related art considers either the single-device scenario, trying to optimise the execution of one or multiple DNNs on constrained devices, or the client-server scenario, which offloads heavy computations to the cloud. Few broaden the view to the possibilities offered by more decentralised infrastructures. We aim to fill this gap via an initial case study applied to a multiview detection system.

In this work, we propose a real-world implementation of a multiview detection system based on the MVDet state-of-the-art detection model~\cite{you2020multiview}. We port the model to C++ using the libtorch~\cite{paszke2019pytorch} and opencv~\cite{bradski2000opencv} libraries and integrate it into a framework for distributed inference. We experiment with it under different computational and networking hypotheses, comparing the traditional client-server approach to a distributed approach, which splits the model execution across multiple devices to test the real benefits of a decentralised infrastructure. Our contributions are:
\begin{itemize}
    \item an open-source implementation of a real-world edge-based system for multiview detection. While our implementation is based on the MVdet model, the methodology can be applied to different models;
    \item extensive experimentation of the proposed system under different computational and network conditions;
    \item a critical discussion on the potentiality of decentralised infrastructure.
\end{itemize}

The paper is structured as follows: Section \ref{sec:intro} briefly introduces the research context, Section \ref{sec:background} introduces the basics for understanding the subsequent sections, Section \ref{sec:methods} details the multiview detection use-case, the implementation methodology and the research choices, Section \ref{sec:experiments} discusses in detail the experimental contribution of this work, and Section \ref{sec:end} wraps up the exposition, summarising the findings and future research directions.

\section{Background}
\label{sec:background}

Edge inference is a research field undergoing strong experimental efforts due to the convergence of many phenomena. The abundance of pervasive computational devices~\cite{Narayanan2020pervasive} offers lots of spare computational power exactly where data are harvested.
Many approaches exploit this latent computing power to process data timely, especially for medical purposes~\cite{rahmani2018exploiting, greco2020trends}.
However, as ML models became increasingly complex and computationally power-consuming, edge devices adapted like in a co-evolution schema~\cite{xu2018scaling}.
Thus, AI-specialised edge devices have flourished, like TPUs~\cite{sengupta2020high}, NPUs~\cite{tan2020fastva}, FPGAs~\cite{gomes2015towards}, and many others based on innovative computing paradigms, like in-memory computing~\cite{hsu2019ai}.
All this hardware-related research effort is devoted to increment the efficiency of AI-related computations, trying to mitigate as much as possible the natural computing and battery life limitation of such devices, but many real-world still incorporate only CPUs as compute due to cost, thermal and power constraints~\cite{iot_19}.
On the other hand, ML practitioners devised new algorithms to exploit edge opportunities.
Some efforts, like model compression~\cite{shah2021model}, pruning~\cite{yu2018nisp}, and distillation~\cite{mullapudi2019online}, focus on reducing as much as possible the model size to improve both its memory and energy impact.
Conversely, model partitioning aims at dividing models into manageable chunks to optimise scheduling on single devices~\cite{CoxBC22} or to offload parts of computation across multiple computing devices according to the capabilities of each one~\cite{benelallam2016efficient, yu2021gillis, pacheco2021calibration}, improving inference latency.
Still, another strategy is to completely avoid using DNNs in favour of simpler, less computationally intensive models such as classical, non-deep ML models.
Despite, the feasibility of this approach in an edge environment~\cite{mittone23mafl}, not all tasks are suited to these models.

ML-based detection systems can be modelled in many ways, but reliance on DNNs is one constant.
Single-view detection~\cite{zhu2014single, ouyang2015partial, zhang2018occlusion, hosang2017learning} addresses the case in which just a single image of a scenario is available for inference at a time.
Such methods can be anchor-based or not and can handle occlusion issues by exploiting techniques such as part-recognition, non-maximal suppression, and repulsion loss.
Additional information exploited by the inference can be domain-specific, like the detection of heads and feet when searching for people, or can be obtained through the use of particular image acquisition tools, such as RGB-D cameras and LIDAR detectors, to obtain single images correlated with more spacial information.
Multiview detection exploits multiple sources of image acquisition to produce a single inference~\cite{you2020multiview, fleuret2007multicamera, su2015multi} to overcome occlusion problems.
The principal research focus in this scenario is aggregating the information retrieved from multiple data sources.
Popular approaches target combining multiple single-view detection, aggregating the features extracted from each image, and applying geometrical transformation to the camera outputs.


\section{Methodology}
\label{sec:methods}

\begin{figure*}[t]
  \begin{center}
  \includegraphics[scale=0.5]{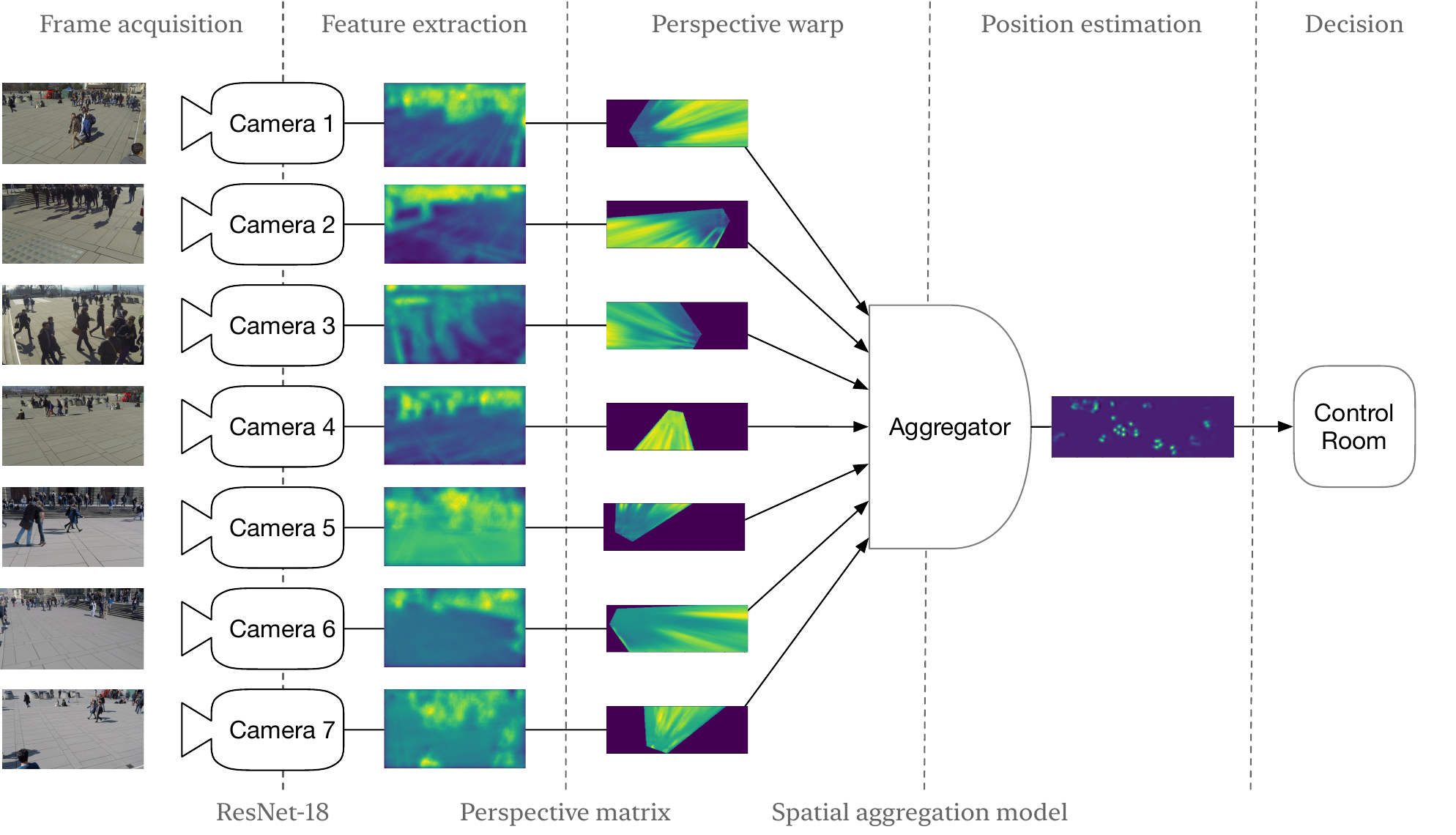}
  \caption{Distributed implementation of MVDet with FastFL. The workflow starts with each camera acquiring the current time step frame; the frame features are then extracted by a ResNet18 and warped according to the camera perspective matrix directly on the edge; then all the warped feature maps are collected by the aggregator, which aggregates them via the spatial aggregation model, producing the position estimation map; this last result is then sent to the control room for operational decision. This process is repeated iteratively.}
  \label{fig:impl}
  \end{center}
\end{figure*}

The MVDet is a state-of-the-art multiview detector that identifies persons standing and moving across an open public area. The main idea is to use a trained base model, e.g. ResNet18 or VGG11, to extract the features from each camera frame and transform it via a homography, i.e. perspective warp, projecting it to the bird eye view of the area. The results are then fed into an aggregation model, which detects the position of all the persons in the area.

Bringing the MVDet model to a real-world edge environment requires methodological and implementational choices.
Starting from the original MVDet code, we identify two different computational stages: a parallelisable one that comprehends the frame acquisition, feature extraction, and perspective warping, and, conversely, a sequential one that is the multiview aggregation.
Since the system is synchronous at the frame level, the multiview aggregation is data-dependant from the previous processing steps: the aggregation can not start until all the camera frames from the current time step have been acquired and processed. 
Given this computational scenario, we propose two different edge implementations of the MVDet system: a more distributed one and a more centralised one.
The distributed implementation is depicted in Figure~\ref{fig:impl}. It takes full advantage of the model partitioning technique, allocating part of the ML models on the edge devices while trying to parallelise the execution as much as possible by assigning all the parallelisable code to a different camera.
Conversely, the centralised implementation takes full advantage of model offloading, allocating all the computational burden on the server while requiring the Camera to acquire the frames.
A more technical discussion on the differences between these two approaches is presented in Section~\ref{subsec:setup}.

Due to the targeted scenario, i.e. edge devices, we stumbled upon the fact that no Distributed ML (DML) framework currently available on the market is designed to be energy-efficient and computationally lightweight. 
The commercial software is Python-based, requiring computational and memory capabilities that not every edge device can afford.
Furthermore, popular DML software is being developed keeping into account mostly user-friendliness and additional security features, such as Homomorphic Encryption and Secure Multiparty Computation, rather than computational performance.
While this strategy comes in handy when dealing with powerful computing devices, an edge-oriented DML framework should care about the amount of resources it consumes (computation, memory, energy), trying to be as efficient as possible.
Also, most commercial DML software are designed to handle in a very solid way only one communication topology, the master-worker one.
Any other communication structure would require heavy software modifications, resulting in a not-as-intended use of the frameworks.
The tree-based inference structure we envision is thus not implementable straightforwardly in current DML software.
Due to these two motivations, efficiency and communication topology malleability, we use an experimental DML framework named FastFederatedLearning (FastFL)~\cite{mittone2023ffl}.
FastFL offers a high-performance C/C++ implementation wrapped with a user-friendly Python interface, allowing user-friendliness and computational performance.
The communication backend is handled by the FastFlow C/C++ high-performance parallel programming library~\cite{aldinucci2017fastflow}, allowing the user to specify custom communication topologies, working both in shared and distributed memory.

The application architecture is based on several building blocks, i.e. specialisations of the \texttt{ff\_node} class, the primary logical unit exposed by FastFlow, connected in a tree-like structure.
Building blocks can be executed in a distributed manner and can communicate by exchanging messages through different backends, like TCP and MPI~\cite{distributedFF}.
At the leaves of the tree, we place multiple \texttt{ff\_monode\_t}, named \textit{Camera}, each representing a different camera in the system. These building blocks are executed in parallel independently. 
For the distributed implementation, a new frame is read at each time step, its features extracted by a pre-trained base model, i.e. ResNet18 in our case, and warped according to a perspective transformation to consider the camera view angle.
Conversely, for the centralised implementation, the Camera just acquires the current time step frame.
Each \textit{Camera} sends the result to the next level of the tree.
This level consists of a \texttt{ff\_minode\_t} called \textit{Aggregator}, which takes multiple inputs and returns a single output.
In the distributed implementation, results collected from all children \textit{Camera} are aggregated by the spatial aggregation model into the final position estimation map.
Conversely, for the centralised implementation, the \textit{Aggregator} node also takes over the feature extraction and perspective warping for each received camera frame.
Finally, the position estimation map is sent to the tree's root, i.e. \texttt{ControlRoom} node, where real-time decisions can be taken in a real-world deployment. Figure~\ref{fig:impl} summarises the distributed implementation architecture.
This procedure is synchronously repeated using an additional \textit{Sync} node connected to all \textit{Camera} (not shown in the figure for simplicity) used to trigger the next frame until all frames are processed.
In case of multiple open areas, the application supports multiple subtrees, one for each area, each consisting of an \textit{Aggregator} with multiple \textit{Camera} as children.

\section{Experimental results}
\label{sec:experiments}

\begin{figure*}[ht]
  \begin{center}
  \includegraphics[width=\textwidth]{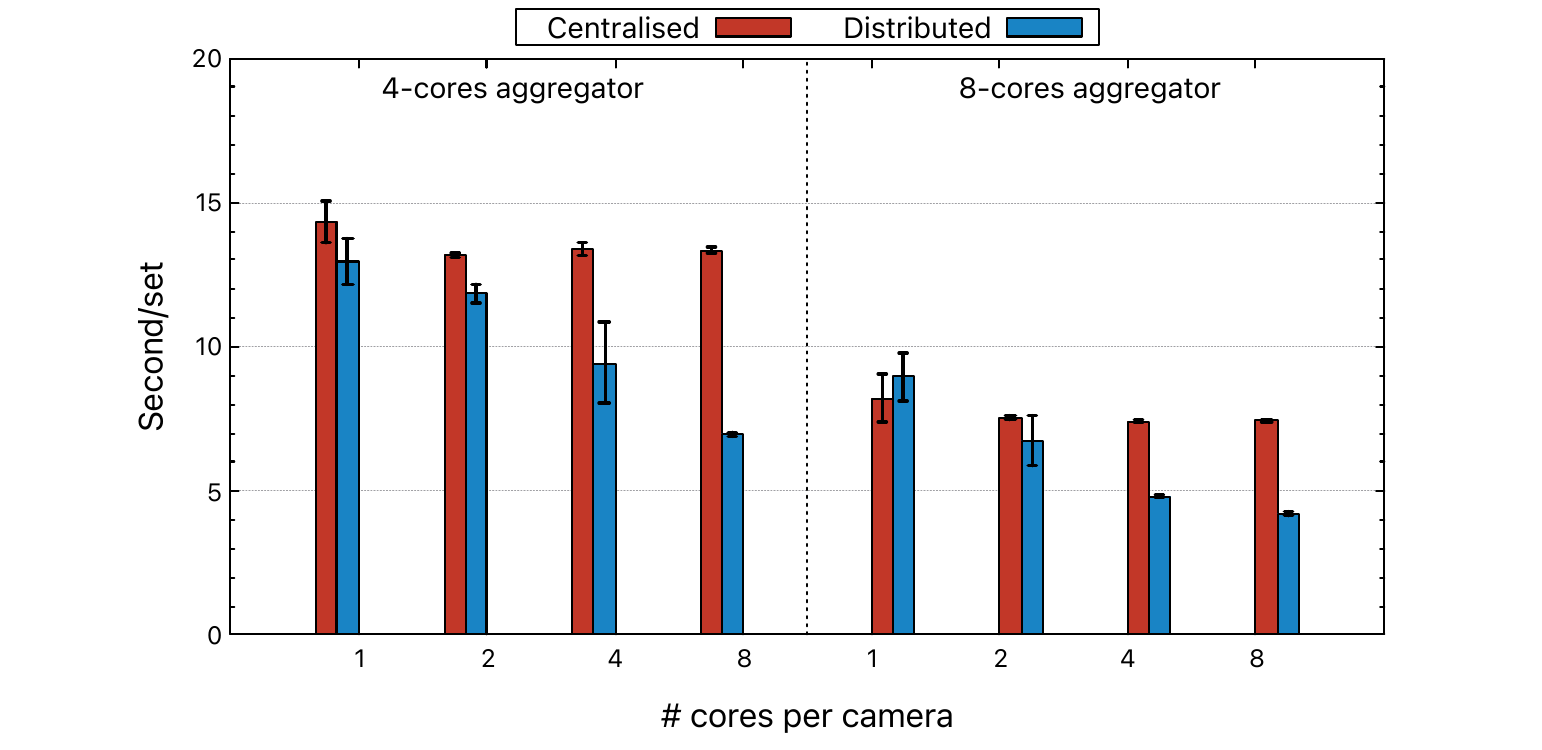}
  \caption{Computational performance of the proposed systems with varying computational power available. This plot compares the computational performance of the distributed MVDet system, measured as seconds taken to process each frame +/- the 95\% confidence interval over 5 runs, in all possible combinations of computational power assigned to the server (4, 8 cores) and cameras (1, 2, 4, 8 cores).}
  \label{fig:times}
  \end{center}
\end{figure*}

\subsection{Setup}
\label{subsec:setup}

{\bf Dataset}: We use the Wildtrack multiview dataset~\cite{wildtrack}.
It comprises 7 static cameras capturing views of a public open area with dense groups of pedestrians standing and walking. 
The dataset provides each camera with the accurate position and view angle and 400 time-synchronised full-HD frames.

{\bf Testbed}:
All experiments were run on the HPC4AI cloud computing facility~\cite{aldinucci2018hpc4ai} exploiting  10 virtual machines, each one equipped with 8 64-bit vCPUs mapping to dedicated cores of an Intel Xeon Gold-6230 @2.10GHz processor (Skylake, IBRS), 16GB RAM, 1 Gb/s interconnection network, and running Ubuntu 22.04 as operating system.
We chose to use CPU-only nodes to better model the capabilities of edge devices, which often forgo GPUs for cost, energy, and thermal constraints. 
In contrast, the single-core performance of modern SoC can be in line with that of a vCPU.
Each system component (7 Camera, 1 Sync, 1 Aggregator and 1 ControlRoom node) is deployed on a different virtual machine.

{\bf Performance metric}: We choose the time needed to process one complete set of camera frames and produce the estimated positions as a performance metric. Experiments are replicated 5 times, and we report the mean plus 95\% confidence interval (CI).

{\bf Deployment scenarios}:
We tested the proposed system under various conditions to emulate different deployment scenarios.
We created two versions of the multiview detection application.
The first is a centralised version, which acts as a baseline representing the traditional client-server approach where each Camera (client) sends the acquired frame to the Aggregator (server). 
In this case, the Aggregator node performs all the model computation, whilst the Camera nodes limit themselves to acquiring the next frame and sending it to the Aggregator.
This implementation implies almost no computational power needed from the Camera but places a heavy computational burden on the Aggregator instead; furthermore, the communications are lighter, since the acquired frames are smaller tensors, size $[3x1920x1080]$, i.e. $6.1$ MB, than the extracted feature tensors, size $[512, 360, 120]$, i.e. $21.6$ MB.
The second is the distributed version, where we split the detection model according to Figure~\ref{fig:impl} leveraging the computing power of each Camera for feature extraction and perspective warping, while leaving to the Aggregator spatial aggregation and final position estimation.

\begin{table*}[tbph]
\caption{Computational performance of the proposed systems with varying computational power and network bandwidth. The first two scenarios adopt a bandwidth comparable with current 5G performance in Italian cities, while the third simulates a resource-constrained edge deployment.} 
\label{tab:network}
    \centering
    {\small
    \begin{tabular}{lllll}
    \toprule
    {\bf \makecell{Aggregator cores}}&{\bf \makecell{Camera bandwidth (up/down)}}&{\bf \makecell{Aggregator bandwidth (up/down)}}&{\bf \makecell{Centralised (s/set)}}&{\bf \makecell{Distributed (s/set)}}\\
    \midrule
    4 cores & 25/284 Mb/s & 1/1 Gb/s & 15.38 & 49.68  \\
    8 cores & 25/284 Mb/s & 1/1 Gb/s & 11.98 & 46.89  \\
    8 cores & 10/10 Mb/s  & 100/100 Mb/s  & 14.22 & 108.79 \\
    \bottomrule
        \end{tabular}
    }       
\end{table*}

{\bf Resource modulation}:
We test the proposed multiview detection system under different combinations of computational power assigned to the critical software components, i.e. Camera and Aggregator, thus simulating how different edge devices with varying performance impact the system.
We modulate the computing power by varying the number of cores available to the different components using taskset and providing hints to the number of threads to spawn to libtorch via the \texttt{MKL\_NUM\_THREADS} and \texttt{OMP\_NUM\_THREADS} environment variables.
Specifically, we tested the system assigning to each Camera 1, 2, 4, or 8 cores, and to the Aggregator 4 or 8 cores, for 16 different computational power configurations.
We also emulate different network conditions by limiting the bandwidth available to the different nodes to study how it affects the performance of the proposed system.
Indeed, edge devices typically need to rely on slower networks, e.g. cellular connections.
Kollaps is a decentralised container-based network emulator, exploiting Docker Swarm (or Kubernetes) to deploy and run distributed computations with specific network topology made of bridges and links with specific upload/download bandwidth, latency, and jitter characteristics enforced using the \texttt{traffic control} capability of the Linux kernel.
In the following experiments, we study the impact of varying upload/download bandwidths of the Camera and Aggregator.
We test the system with 10/10 Mbps per Camera and 100/100 Mbps for the Aggregator, representing a low bandwidth scenario. We then increase the Camera bandwidth to 25/284 Mb/s, which represents the average upload/download speeds achieved by the best 5G network in Italy in 2022~\cite{opensignal}, and consequently increase the network bandwidth of the aggregator to 1/1 Gb/s to cope with the faster aggregated bandwidth from the Cameras. The latter configuration is tested assigning 4 or 8 cores to the Aggregator. 


\subsection{Results}

Figure~\ref{fig:times} presents the results across the 16 computational power configurations.
One can notice how the centralised implementation performance is basically unaltered by the amount of computing power given to the Camera, while doubling the number of cores given to the Aggregator almost halves the amount of time required to process a single frame, i.e. from 13.57 s to 7.66 s on average.
Conversely, the distributed approach is more sensible to the computing power given to both Camera and Aggregator, steadily increasing its computational performance when the computing resources given to the system (Camera and Aggregator nodes) increase from $12.97$ s/set using a total of $7x1+4 = 11$ vCPUs to $4.23$ s/set using a total of $7x8+8 = 64$ vCPUs.

By comparing the two systems, it is clear that the distributed implementation obtains globally lower computational times, achieving better computational performance with respect to the centralised approach thanks to exploiting the computational power spread over the computing continuum. The higher the computing power of the Camera, the more significant the performance gap, i.e. up to 1.92x faster with 4 cores per Aggregator and 8 cores per Camera.
However, also the network plays a role. Indeed, in the 1~(8) cores per Camera~(Aggregator), the centralised approach beats the decentralised approach by a small margin due to the increased communication time to transmit the 3.5x larger feature maps instead of the plain captured frames. 

Looking at Table~\ref{tab:network}, it is possible to observe how both implemented systems behave under different network bandwidth conditions.
To better simulate a low-power edge system, each camera node is allocated 2 cores.
As can be seen, the distributed implementation particularly suffers from bandwidth limits.
As explained before, the feature maps are way heavier than the plain frames; hence, bandwidth limits have a particular impact on the system performance, regardless of the amount of computing power allocated to each component.
This behaviour can be noted especially in the last row of Table~\ref{tab:network}, in which the centralised system is 7.65x times faster than the distributed one.
In a 5G network, which has higher bandwidths, this speedup decreases to 3.23x-3.91x, a reduction of 42.22\%-51.11\% with respect to the previous scenario.
Instead, when the network is not the bottleneck,
the distributed approach is the best in almost all scenarios, as shown in Figure~\ref{fig:times}. Please note that here we ported the MVDet model as is without any optimisation for communication not to alter the model performance. While possibly having a detrimental effect on the model performance, the communication cost could be lowered by compressing the feature maps or retraining the model with feature maps having a lower number of channels. 

This comparison clearly shows how the environment can influence the real-world deployment of an edge inference system.
Computational power and network bandwidth allocated to each computing continuum element are crucial, but there is no one-fits-all strategy to implement such systems.
Workload distribution across the continuum can often become disadvantageous if it leverages critical resources that the centralised counterpart, instead, is not so reliant on.
Nevertheless, the same distributed deployment can efficiently exploit all the available resources in an ideal scenario, thus outperforming the more centralised approach.
Since environmental conditions change over time, future edge inference systems should consider this variability and try to accommodate and adapt to it, especially if they claim to be flexible, reliable, and efficient.

\section{Conclusions}
\label{sec:end}

This research work presented a real-world, edge-inference multiview detection system implemented with the FastFL framework.
We showed how to move from an offline ML model to a high-performance, distributed inference system, measuring how different computing and network conditions can affect a real-world implementation.
We showed how the environmental setting can affect the real-world performance of the computation implemented according to two different strategies: the centralisation of the computation on more powerful devices and the distribution of the computation to exploit as much edge computing power as possible.

We demonstrated that no strategy is inherently better than the other; however, they offer different properties that can offer better performance in different scenarios, pushing for the need for dynamic, adaptable edge inference systems. 
Implementing more sophisticated ML techniques in the proposed software, such as model distillation, quantisation and compression, and the exploitation of specialised hardware, such as TPUs or NPUs, could compensate for some of the disadvantages of the distributed system and are thus worth investigating as future work.

\begin{acks}
This work receives EuroHPC-JU funding under grant no. 101034126, with support from the Horizon2020 programme (the European PILOT). This work is also supported by the Spoke "FutureHPC \& BigData” of the ICSC – Centro Nazionale di Ricerca in "High Performance Computing, Big Data and Quantum Computing", funded by European Union – NextGenerationEU.
\end{acks}

\bibliographystyle{ACM-Reference-Format}
\bibliography{bibs/bibliography}

\end{document}